\documentclass[prd,superscriptaddress,preprint]{revtex4}

\usepackage{mathrsfs,amsmath}
\usepackage{eurosym}
\usepackage{amssymb}
\usepackage{bbold}

\usepackage{latexsym,epsfig,epstopdf}
\usepackage{amssymb,amsmath,amsthm}

\usepackage{times}
\usepackage{color}

\usepackage{lmodern}
\usepackage[utf8]{inputenc}

\usepackage{graphicx}

\usepackage{color}

\usepackage{braket}

\usepackage[mathscr]{euscript} 

\def\be{\begin{equation}}
\def\ee{\end{equation}}
\def\bea{\begin{eqnarray}}
\def\eea{\end{eqnarray}}

\begin{document}

\title{Zero modes of velocity field and topological invariant in quantum torus}
\author{Annan Fan}
\affiliation{School of Physics, Sun Yat-Sen University, Guangzhou, 510275, China}
\author{Shi-Dong Liang}
\altaffiliation{Email: stslsd@mail.sysu.edu.cn}
\affiliation{School of Physics, Sun Yat-Sen University, Guangzhou, 510275, China}

\date{\today }
\begin{abstract}
We propose the velocity field approach to characterize topological invariants of quantum states. We introduce the indexes of the velocity field flow based on the zero modes of the velocity field and find that these zero modes play the role of effective topological charges or defects linking to Euler characteristic by the Poincar\'{e}-Hopf theorem.
The global property of the indexes is topological invariants against the parameter deformation. We demonstrate this approach by the quantum torus model and compare the topological invariant with that obtained from the Chern number. We find that the physical mechanism of the topological invariant based on the zero modes of the velocity field is different from that of the topological invariant by the Chern number. The topological invariant characterized by the velocity field describes a homeomorphic topological invariant associated with the zero modes on the submanifold of the base manifold of the SU(2)-fibre bundle for quantum torus, whereas the Chern number characterizes a homotopy invariant associated with the exceptional points in the Brillouin zone. We also propose the generalized winding number in terms of the velocity field for both Hermitian and non-Hermitian systems. This gives a connection between the zero mode and winding number in the velocity space.
These results enrich the topological invariants of quantum states and promises us a novel insight to understanding topological invariants of quantum states as well as expected to be further applied in more generic models.
\end{abstract}

\maketitle



\section{Introduction} 
Topological invariants of quantum states
have attracted growing attention due to their novel fundamental issues and potential applications in condensed matter physics and quantum computation.\cite{Ramy,Gong,Kohei}
In particular, a family of quantum Hall effects was discovered to connect with geometric and topological properties of quantum states described by the Berry phase, Berry curvature, winding number and Chern number, which inspires a lot of attempts to explore what connection between novel physical phenomena and their mathematical structures.\cite{Chiu,Kane1,Qi,Chruscinski,Annan}

Early studies were focused on the geometric properties of the wave function of quantum states, such as the Berry phase, and Berry curvature as a gauge potential and gauge field associated with the geometric and topological properties of quantum states.\cite{Chiu,Chruscinski,Annan,Bohm,Ghatak,Liang} Along this direction, one found that topological invariants of quantum states can be characterized by some topological indexes, such as winding number and Chern number.\cite{Chiu,Qi,Ghatak,Zhang} Particularly, the boundary states play a crucial role to dominate topological invariant as a bulk-boundary correspondence. \cite{Chiu,Zhang,Lee,Yao} Another issue is the classification of the topological equivalent classes based on the symmetry of systems. It has been found that there are 10-fold topological equivalent classes for Hermitian systems based on the Altland-Zirnbauer(AZ)symmetry classification and 38-fold topological equivalent classes for non-Hermitian systems due to the additional sublattice symmetry and pseudo-Hermticity of non-Hermitian systems. \cite{Gong,Kohei}

One discovered that topological invariants involve two crucial features of quantum states, the boundary states and the energy band gap.
The non trivial topological invariant depends on the features of the boundary states near the Fermi energy in the gapped modes .\cite{Chiu,Lee,Yao,Yu,Yin}
The energy band gap based on the symmetries of systems protects topological invariants of quantum states against the parameter deformation.\cite{Chiu,Ghatak,Zhang}

Recently, one proposed the vorticity concept defined by the complex angle of the complex energy band to characterize topological invariant in non-Hermitian systems, \cite{Ghatak,Fu} which has been shown to be equivalent to the winding number classifying the topological invariants for Hermitian systems, but in the complex energy space. \cite{Kohei,Fu} In particular, the topological defects and gapless modes in insulators and superconductors were discovered to be related to a generalized bulk-boundary correspondence,\cite{Kane2} which provides an approach to classify temporal pumping cycles, such as the Thouless charge pump, fermion parity pump as well as Majorana zero modes. \cite{Kane2}
Along this direction, one introduced a planar vector field to construct a generalized winding number and linking number to characterize topological invariants of exceptional points for non-Hermitian systems.\cite{Song}
These results inspire us to propose a novel approach to explore topological invariant of quantum states based on the zero modes of the velocity field induced by the energy band structures.

In this paper, we will propose the velocity field approach  as a vector field on the sub manifold of the base manifold of the SU(2)-fibre bundle to explore the topological invariants of quantum states. In Sec. II, we will introduce the velocity field of the Bloch electrons and define the index of the velocity field based on the stationary points or zero mode (nodes) of the velocity field. We find that these zero modes of the velocity field can be viewed as effective topological charges or defects associated with the topological invariant. The index of the velocity field based on the zero modes is related to the Euler characteristic by the Poincar\'{e}-Hopf theorem,\cite{Eber}. In Sec. III, we will demonstrate this velocity field approach by quantum torus model and yield the topological invariants of quantum states characterized by the Euler characteristic, which is consistent with the standard mathematical result. Finally in Sec. IV, we will discuss the relationship between this topological invariant characterized by the velocity field flow and that by the Chern number, and we will give the conclusions.

\section{Velocity field in quantum systems}
\textit{Velocity field }:
Let us consider a quantum system described by a bound Hermitian Hamiltonian in the Brillouin zone (BZ) with a parameter space, $H(\mathbf{k},\lambda)$, where $\mathbf{k}\in BZ^{d}$ is the $d$-dimensional BZ and $\lambda\in \mathcal{M}^{p}_{\lambda}$ is a set of parameters in the $p$-dimensional parameter space.
The eigen equations is given by
\begin{equation}\label{HH}
H(\mathbf{k},\lambda)|\psi_{n}(\mathbf{k},\lambda)\rangle=E_{n}(\mathbf{k},\lambda)|\psi_{n}(\mathbf{k},\lambda)\rangle
\end{equation}
Suppose that the Hilbert space for this system is separable, the eigen vectors of the Hamiltonian are consisted of
an orthogonal basis on the Hilbert space, $\langle\psi_{m}(\mathbf{k},\lambda)|\psi_{n}(\mathbf{k},\lambda)\rangle=\delta_{mn}$ and
the complete relation $\sum_{n}|\psi_{n}(\mathbf{k},\lambda)\left\rangle\right\langle\psi_{n}(\mathbf{k},\lambda)|=I$,
where $I$ is the identity matrix. The eigen vectors describe the wave function of the Bloch electrons and their corresponding eigen energies describe the energy bands. In principle, all physical properties of quantum states depend on the wave functions and their corresponding energy bands.
In particular, the electronic transport in condensed matter physics, such as the electric, heat conductances and  Hall conductance, depend on the velocity of the Bloch electrons. The velocity of the Bloch electron is defined  by\cite{Mermin}
\begin{equation}\label{VF0}
\mathbf{v}_{n}(\mathbf{k},\lambda)=\frac{1}{\hbar}\nabla_{\mathbf{k}}E_{n}
\end{equation}
where $\nabla_{\mathbf{k}}$ is the gradient operator. The velocity of the Bloch electron forms a covector field in the BZ, namely $\mathbf{v}_{n}(\mathbf{k},\lambda)=dE_{n}$, where we set $\hbar=1$ without losing generality in the following section.
We will study the analytic properties of the velocity field in the BZ to reveal some novel physical properties of quantum states.

\textit{Poincar\'{e}-Hopf theorem and topological invariants}:
In order to explore the relationship between the velocity field of the Bloch electrons and the topological invariant, we consider a typical two-level model with the Hamiltonian in the 2-dimensional (2D) BZ given by
\begin{equation}\label{nH2}
H(\mathbf{k},\mathbf{\lambda})=\mathbf{h}(\mathbf{k},\mathbf{\lambda})\cdot \mathbf{\sigma}
\end{equation}
where $\mathbf{\sigma}$ is the Pauli matrix, and $\mathbf{\lambda}\in \mathcal{M}^{p}$ is a set of parameters. The Hamiltonian is Hermitian when $\mathbf{h}(\mathbf{k},\mathbf{\lambda})\in \mathbb{R}$. From the geometric point of views, $\mathbf{h}(\mathbf{k},\mathbf{\lambda})$ plays a role of the submanifold of the base manifold of the SU(2)-fibre bundle (Mathematically, the base manifold of the SU(2)-fibre bundle is $\mathbb{R}^{3}$).\cite{Chiu,Chruscinski}
\begin{equation}\label{TF}
\mathbf{h}(\mathbf{k},\mathbf{\lambda})
=\left\{(h_{x}(\mathbf{k},\mathbf{\lambda}),h_{y}(\mathbf{k},\mathbf{\lambda}),h_{z}(\mathbf{k},\mathbf{\lambda}))|\mathbf{k}\in(0,2\pi)^2\right\},
\end{equation}
which actually can be viewed as a 2D surface of the compact manifold embedded in the 3D manifold for given parameters $\lambda\in\mathcal{M}^{p}$.

In general, the velocity field can be redefined as the vector field\cite{Eber}
\begin{equation}\label{VF2}
\mathbf{v}_{n}(\mathbf{k},\mathbf{\lambda})  = v_{n,x}(\mathbf{k},\mathbf{\lambda})\frac{\partial \mathbf{h}}{\partial k_x}
+v_{n,y}(\mathbf{k},\mathbf{\lambda})\frac{\partial \mathbf{h}}{\partial k_y}
\end{equation}
on the submanifold of the base manifold $\mathbf{h}(\mathbf{k},\mathbf{\lambda})$ for given parameters $\mathbf{\lambda}$,
The components of the velocity field in (\ref{VF2}) are given by
\begin{equation}
v_{n,x}(\mathbf{k},\mathbf{\lambda}) = \frac{\partial E_n}{\partial k_x},  \quad  v_{n,y}(\mathbf{k},\mathbf{\lambda})=\frac{\partial E_n}{\partial k_y}
\end{equation}
and
\begin{subequations}\label{CBMf1}
\begin{eqnarray}
\frac{\partial\mathbf{h}}{\partial k_x} &=& \left(\frac{\partial h_{x}}{\partial k_x},\frac{\partial h_{y}}{\partial k_x},\frac{\partial h_{z}}{\partial k_x}\right) \\
\frac{\partial\mathbf{h}}{\partial k_y} &=& \left(\frac{\partial h_{x}}{\partial k_y},\frac{\partial h_{y}}{\partial k_y},\frac{\partial h_{z}}{\partial k_y}\right)
\end{eqnarray}
\end{subequations}
are the basis in the submanifold of the base manifold,

Note that the velocity field in the BZ is the local representation of the vector field on the submanifold of the base manifold of the fibre bundle. In other words, we can capture the topological invariant of the submanifold of the base manifold from the analytic behaviors of the velocity field in the BZ when the basis in (\ref{CBMf1}) on the submanifold of the base manifold is compatible with the basis in the BZ, namely they are linear independent and can be expressed as
\begin{equation}\label{CTJD1}
\frac{\partial\mathbf{h}}{\partial k_x}\times \frac{\partial\mathbf{h}}{\partial k_y}
\neq 0.
\end{equation}

Due to the particle-hole symmetry of the quantum torus Hamiltonian, the velocity field is independent of the energy band indexes. Thus, we ignore the band index in the following section. When the velocity field contains some finite isolated regular zero modes $\mathbf{k}_0$,  namely zero modes or nodes $\mathbf{v}(\mathbf{k}_0)=0$,
we can define the index of these zero modes,\cite{Eber}
\begin{equation}\label{Ind1}
I_{d} \left[\mathbf{v'}(\mathbf{k}_0,\lambda)\right]:=(-1)^{S_{g}} .
\end{equation}
where $S_{g}$ depends on the determinant of the velocity field,
\begin{equation}\label{JDVF1}
\det \left[\mathbf{v'}(\mathbf{k}_0,\lambda)\right]:=
\begin{vmatrix}
\frac{\partial v_{x}(\mathbf{k},\lambda)}{\partial k_{x}} & \frac{\partial v_{x}(\mathbf{k},\lambda)}{\partial k_{y}} \\
\frac{\partial v_{y}(\mathbf{k},\lambda)}{\partial k_{x}} & \frac{\partial v_{y}(\mathbf{k},\lambda)}{\partial k_{y}}
\end{vmatrix}_{\mathbf{k}=\mathbf{k}_{0}}
\end{equation}
and $S_{g}=0$ for $\det\left[\mathbf{v'}(\mathbf{k}_0,\lambda)\right]>0$ and $S_{g}=1$ for $\det\left[\mathbf{v'}(\mathbf{k}_0,\lambda)\right]<0$.
In other words, $S_g$ means the sign of the determinant of the velocity field at the zero mode. \cite{Eber} Namely,
$I_{d} \left[\mathbf{v'}(\mathbf{k}_0,\lambda)\right]=1$ for the sink or source, whereas $I_{d} \left[\mathbf{v'}(\mathbf{k}_0,\lambda)\right]=-1$ for the saddle point of the zero modes.\cite{Eber}

Mathematically, the Poincar\'{e}-Hoft theorem provides a description of the topological invariant of the velocity field associated with the Euler characteristic of the manifold. The Poincar\'{e}-Hoft theorem tells us that the velocity field $\mathbf{v}$ as a vector field on the manifold (\ref{TF}) is continuous and contains finite zero modes, when the manifold is compact and orientable, the sum of indexes of these zero modes is the Euler characteristic of the manifold,\cite{Eber}
\begin{equation}\label{PHT}
\sum_{j}I_{d} \left[\mathbf{v}(\mathbf{k}_{0,j},\lambda)\right]=\chi(\lambda),
\end{equation}
where the sum $j$ covers all the zero modes in the BZ.
Thus, the global properties of the indexes of the velocity field in the BZ yield the topological invariant on the submanifold of the base manifold characterized by the Euler characteristic. This topological invariant depends on the parameters, $\lambda\in\mathcal{M}^p$ such that we can obtain the topological phase diagram in the parameter space.

In general, the topological invariant of the velocity field in the BZ can characterize the topological invariant of the submanifold of the base manifold of quantum systems based on the conditions of the Poincar\'{e}-Hoft theorem,\cite{Eber} which are
(1) the submanifold of the base manifold in (\ref{TF}) is continuous to $C^\ell$ with $\ell\geq 2$, compact and orientable (non-degenerate), (2) the basis in (\ref{CBMf1}) is compatible with the basis in the BZ, namely linear independent, and satisfies (\ref{CTJD1}), and (3) the zero modes as the effective topological charges are isolated and finite in the BZ. When these conditions are fulfill for any physical system, the topological invariant of the velocity field in the BZ can characterize the topological invariant of the submanifold of the base manifold of quantum systems by the Euler characteristic. Mathematically, the covector of the velocity field in the BZ is a
pull-back of the velocity field on the submanifold.

In addition, this velocity field approach can be generalized to higher dimensional systems because the Poincar\'{e}-Hoft theorem are available for the finite dimensional manifold.

\section{Topological invariants in quantum torus }
\textit{Quantum torus model}:
As a typical example of two level systems, let us consider a quantum torus model. The Hamiltonian can be given by
\begin{equation}\label{QT1}
H(\mathbf{k},\mathbf{\lambda})=\mathbf{h}(\mathbf{k},\mathbf{\lambda})\cdot \mathbf{\sigma}
\end{equation}
where
\begin{subequations}\label{hhh2}
\begin{eqnarray}
h_{x}(\mathbf{k},\mathbf{\lambda}) &=& r_{0}\cos k_{x}+c,\\
h_{y}(\mathbf{k},\mathbf{\lambda}) &=& r_{0}\sin k_{x}\\
h_{z}(\mathbf{k},\mathbf{\lambda}) &=& r\sin k_{y},
\end{eqnarray}
\end{subequations}
where $r_0\equiv\sqrt{r^{2}\sin^{2}k_y+(R+r\cos k_y)^2}$. The parameter $R$ is the radius of the torus and $r$ is the radius of the ring, $R>r$. $c$ is a constant for shifting the $h_x$-axis of the torus ($c>0$) for exhibiting more explicitly the velocity field flow on the torus. The wave vector is within the 2D BZ, $0\leq k_x,k_y\leq 2\pi$. The energy bands of the model are obtained by
\begin{equation}\label{EQT1}
E_{\pm}=\pm\sqrt{h^{2}_x+h^{2}_y+h^{2}_z}.
\end{equation}
The velocity field in the BZ is obtained by
\begin{subequations}\label{vfqt}
\begin{eqnarray}
v_{x}(\mathbf{k},\mathbf{\lambda}) &=& \widehat{\mathbf{h}}\cdot \frac{\partial \mathbf{h}}{\partial k_x},\\
v_{y}(\mathbf{k},\mathbf{\lambda}) &=& \widehat{\mathbf{h}}\cdot \frac{\partial \mathbf{h}}{\partial k_y}
\end{eqnarray}
\end{subequations}
where $\widehat{\mathbf{h}}=\frac{\mathbf{h}}{h}$ is the unit vector of $\mathbf{h}$ and $h=\sqrt{h^{2}_x+h^{2}_y+h^{2}_z}$. We ignore the band index $\pm$ of the velocity field without losing generality.  The local representation of the velocity field on the manifold of quantum torus is given by
\begin{equation}\label{VF3}
\mathbf{v}(\mathbf{k},\mathbf{\lambda})  =  v_x(\mathbf{k},\mathbf{\lambda})\frac{\partial \mathbf{h}}{\partial k_x}
+v_y(\mathbf{k},\mathbf{\lambda})\frac{\partial \mathbf{h}}{\partial k_y}
\end{equation}
It is easy to check $\frac{\partial \mathbf{h}}{\partial k_x}\cdot\frac{\partial \mathbf{h}}{\partial k_y}=0$, which implies the basis on the manifold are orthogonal and the basis transformation between the BZ to the submanifold of the base manifold $\left(\frac{\partial \mathbf{k}}{\partial k_{x}},\frac{\partial \mathbf{k}}{\partial k_{y}}\right)\rightarrow\left(\frac{\partial \mathbf{h}}{\partial k_{x}},\frac{\partial \mathbf{h}}{\partial k_{y}}\right)$ is conformal.
Note that the submanifold of the base manifold is a typical torus $T^{2}$, which is compact and orientable and the velocity field is continuous. Thus, the topological invariant of velocity field on the submanifold of the base manifold (torus) can be characterized by the Euler characteristic based on the Poincar\'{e}-Hopf theorem.

\begin{figure}[htbp]
	\centering
	\includegraphics[width=1\linewidth]{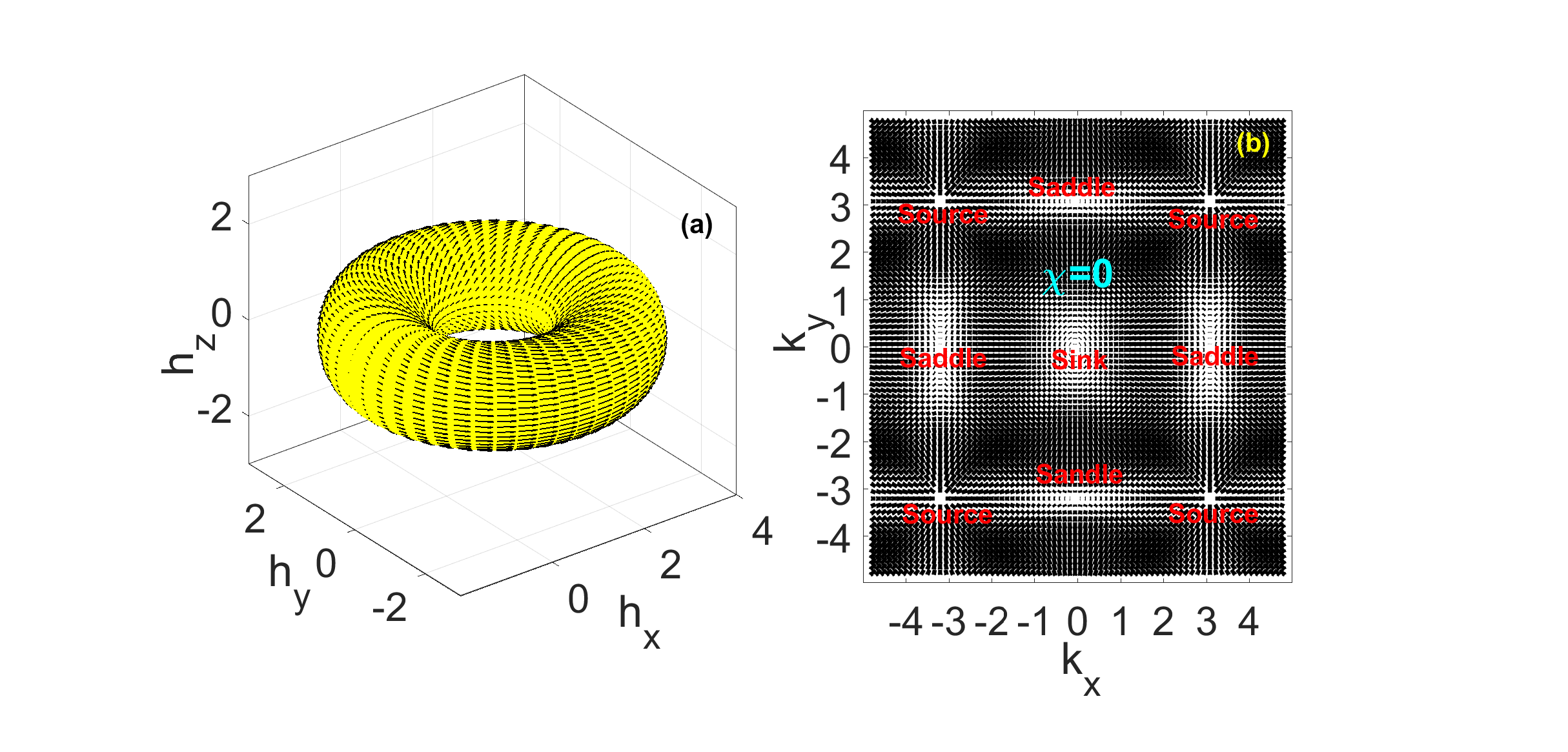}
	\caption{ Online Color: (a) The velocity field on the quantum torus. (b) The velocity field in the BZ, where $R=3$, $r=1$ and c=1.}
	\label{fig1}
\end{figure}

Thus, we can investigate the features of the velocity field and explore their connections to topological invariants based on the Poincar\'{e}-Hopf theorem.\cite{Eber}
By using the energy band in (\ref{hhh2}) and (\ref{EQT1}), the velocity field is obtained by
\begin{subequations}\label{VFxy2}
\begin{eqnarray}
v_{x}(\mathbf{k},\mathbf{\lambda}) &=& -\frac{r_{0}c\sin k_x}{h},\\
v_{y}(\mathbf{k},\mathbf{\lambda}) &=& -\frac{rR}{h}(1+\frac{c}{r_0}\cos k_{x}-\frac{r}{R}\cos k_y)\sin k_y.
\end{eqnarray}
\end{subequations}
The zero modes mean $v_{x}(\mathbf{k},\mathbf{\lambda}) =v_{y}(\mathbf{k},\mathbf{\lambda}) =0$. It is easy to check the zero modes emerge at $k_x=0,\pm \pi$ and $k_y=0,\pm \pi$.

\textit{Topological invariants}: The zero modes of the velocity field play the effective topological charges or defects associated with the topological invariant of the torus, which is characterized by the Euler characteristic based on the Poincar\'{e}-Hopf theorem. We numerically plot the velocity field flow in the BZ and find out the zero modes and their indexes, which yields the Euler characteristic of the submanifold of the base manifold.

It should be noted that when the zero mode of the velocity field emerges at the boundary of the first BZ, the index of the velocity field is counted $\pm\frac{1}{2}$ of the index because they share two BZs. When the zero mode of the velocity field emerges at the corners of the first BZ, the index of the velocity field is counted $\pm\frac{1}{4}$ of the index because they share four BZs. Note that the particle-hole symmetry of the quantum torus model, we investigate only the velocity field of the lower energy band without losing generality.

Fig.\ref{fig1} shows the velocity fields of the quantum torus, in which Fig.\ref{fig1}(a) the velocity field on the torus and (b) is the velocity field in the BZ. We can see that there is a sink at the original point $(0,0)$, a source at every corner $(\pm \pi, \pm\pi)$ and a saddle point at every boundary, $(0,\pm\pi)$ and $(\pm\pi,0)$ of the BZ. The total index is obtained
\begin{eqnarray}\label{Ind1}
\chi &=& (-1)^{S_{sink}}+4\frac{1}{4}(-1)^{S_{source}}+4\frac{1}{2}(-1)^{S_{saddle}} \nonumber\\
   &=& 1+1-2=0,
\end{eqnarray}
which is consistent with the Poincar\'{e}-Hopf theorem applied to the torus, in which $\chi=2-2g$,\cite{Eber} where $g$ is the genus of the manifold. The genus of torus is $g=1$, namely $\chi=0$.

\begin{figure}[htbp]
	\centering
	\includegraphics[width=1.3\linewidth]{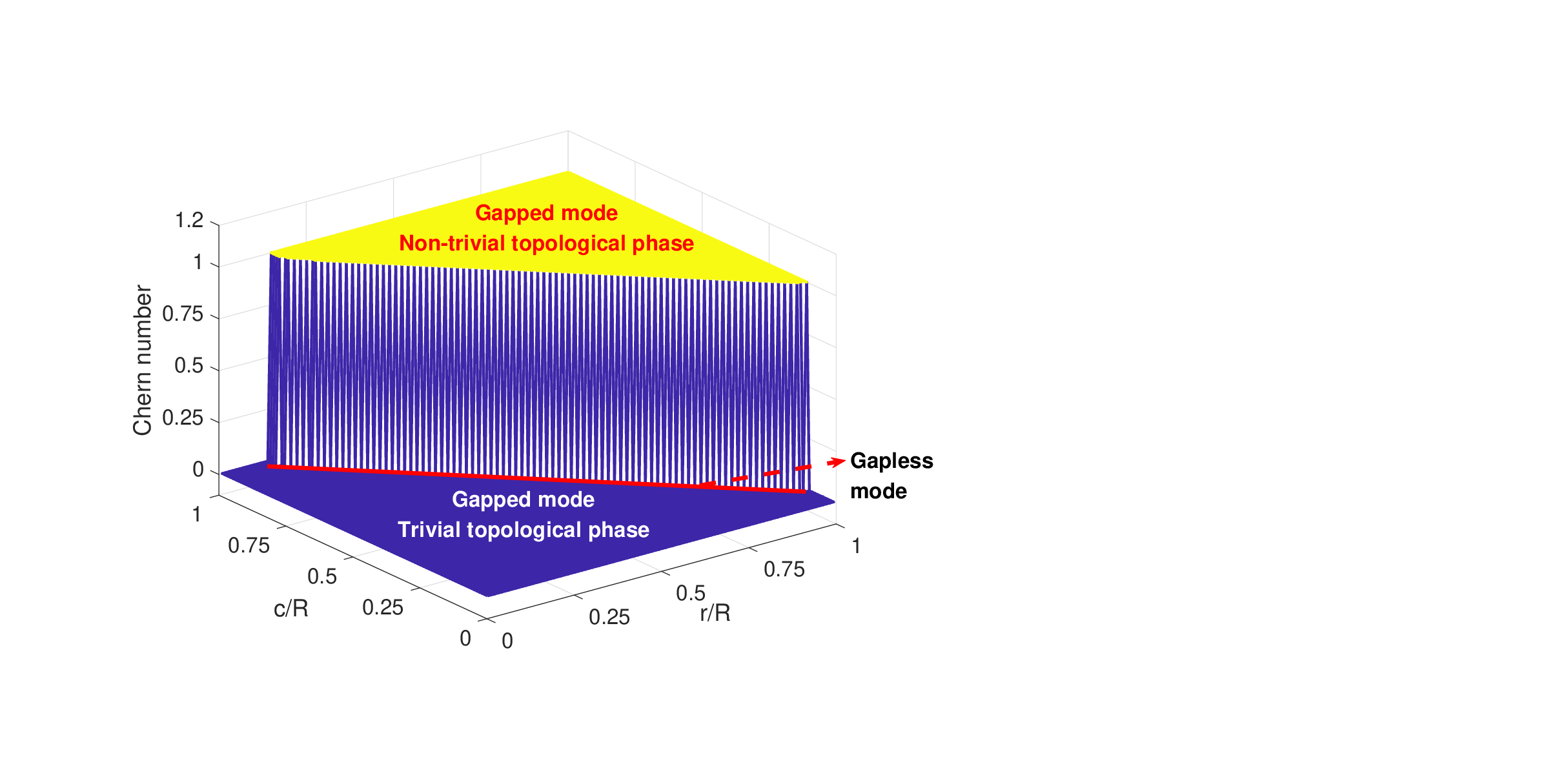}
	\caption{ Online Color: The Chern number $C$ of the quantum torus in the parameter space. The yellow region is $C=1$ for the non-trivial topological phase and the blue region is $C=0$ for the trivial topological phase of the gapped modes. The red line is the gapless modes in the parameter space.}
	\label{fig2}
\end{figure}

On the other hand, the topological invariant of quantum torus can be characterized by the Chern number,\cite{Chiu,Annan}
\begin{equation}\label{CNqt1}
C= \frac{1}{4\pi}\int_{\textrm{BZ}}\mathbf{\widehat{h}}\cdot
\left(\frac{\partial \mathbf{\widehat{h}}}{\partial k_x}\times\frac{\partial \mathbf{\widehat{h}}}{\partial k_y}\right)d^{2}\mathbf{k}
\end{equation}
where $\mathbf{\widehat{h}}$ is the unit vector of $\mathbf{h}$. The energy band gap closes at $h=0$.
Fig.\ref{fig2} shows numerically the Chern number of the quantum torus by (\ref{CNqt1}). The Chern number $C=1$ characterizes the non-trivial topological phase of the gapped mode in the parameter space and $C=0$ represents the trivial topological phase of the gapped mode. The phase transition line (red line) between the trivial and non-trivial phases is the gapless mode in the parameter space. It can be seen that the Chern number depends on the parameter $c$, which is different from the Euler characteristic $\chi$ by the velocity field approach.

\section{Generalized winding number in velocity space}
To compare the physics of the velocity field with the Chern number and winding number, we define a generalized winding number in the velocity space,
\begin{equation}\label{VFWN1}
w_{h}=\frac{1}{2\pi}\oint_{C}d\mathbf{k}\cdot\left(\widehat{v}_{x}\nabla_{\mathbf{k}}\widehat{v}_{y}-\widehat{v}_{y}\nabla_{\mathbf{k}}\widehat{v}_{x}\right),
\end{equation}
where $\widehat{v}_{\alpha}=\frac{\widehat{v}_{\alpha}}{|v|}$ with $\alpha=x,y$. The integral loop $C$ is the velocity evolution in the velocity space. The winding number $w=1,0$ for the zero mode inside and outside of the integral loop in the velocity space, respectively.

For non-Hermitian system, the energy bands are complex in general, $E_{n}=E_{n}^{R}+iE_{n}^{I}$, we can also introduce a complex velocity field by $v=v^R+iv^I$, where $ v^R=\nabla E^{R}$ and  $v^I=\nabla E^{I}$ that neglect the subscript of the energy bands and $\hbar$ without lose of generality. The generalized winding number is defined by\cite{Song}
\begin{equation}\label{VFWN2}
w_{nh}=\frac{1}{2\pi}\oint_{C}d\mathbf{k}\cdot\left(\widehat{v}^{R}\nabla_{\mathbf{k}}\widehat{v}^{I}-\widehat{v}^{I}\nabla_{\mathbf{k}}\widehat{v}^{R}\right),
\end{equation}
where $\widehat{v}^{\alpha}=\frac{\widehat{v}^{\alpha}}{|v|}$ with $\alpha=R,I$. Similarly, the integral loop $C$ is the complex velocity evolution in the complex velocity plane. The winding number $w=1,0$ for the zero mode inside and outside of the integral loop in the velocity space, respectively.

These generalized winding numbers also characterize topological invariants based on the zero modes of the velocity field beyond the conditions of the Poincar\'{e}-Hopf theorem. Interestingly, the generalized winding number for Hermitian systems is written in terms of the angle, $\theta=\arctan \frac{v_y}{v_x}$, and those of the non-Hermitian system is written in terms of the angle, $\theta=\arctan \frac{v^I}{v^R}$. In other words, the generalized winding number describes the interplay between the real and imaginary parts of the velocity field for non-Hermitian systems. The quantum state is robust against the parameter deformation in the topological phase even though there exists the dissipative effects for non-Hermitian systems. However, what physics behind the generalized winding number is a worth studying issue further.

In general, the density of states depend on the velocity field. Many physical properties, such as the electronic and heat conductances, depend on the density of states near the Fermi energy. The zero modes of the velocity field are the singularity points called as the von Hove singularities,\cite{Mermin} which correspond to the divergence of the local density of states. The zero modes play a role of the effective topological charge or defect, dominating the topological invariant of the quantum states. The electronic current is proportional to the velocity of the Bloch electrons. Consequently, the topological invariant based on the velocity field plays a crucial role in many physical properties and potential applications.

\section{Conclusions and outlooks}

What are differences between the topological invariants based on the velocity field, the Chern number and the winding number?
The topological invariant based on the velocity field depends on the zero modes and its flow near the zero modes.
The zero modes means that the Bloch electrons stop at some points in the BZ or on the surface of the manifold. These zero modes can be regarded as effective topological charges or defects,\cite{Kane2} which provide a novel way to classify topological invariants of quantum states.
The global property of the zero modes defined by the total index of the velocity field connects to the Euler characteristic based on the Poincar\'{e}-Hopf theorem. This gives a homeomorphic invariant of the subbase manifold of the SU(2)-fibre bundle for quantum torus. In other words, the velocity field approach can reveal the topological invariant of the quantum states in terms of the geometric object from a mathematica point of view. However,
The connection with the Euler characteristic depends on the basis transformation between the BZ and the submanifold whether compatible each other and the conditions of the Poincar\'{e}-Hopf theorem whether are satisfied. In other words, this connection is no longer hold for the noncompact or nonorientable base manifold.

The topological invariant characterized by the Chern number and winding number depends on the exceptional points inside or outside of the 2D surface of the submanifold. The exceptional point means the closure of the energy band. The topological invariant is against the deformation of the submanifold led by varying parameters as a homotopic map. Due to the Chern number is not well-defined at the exceptional points, as the parameters vary the exceptional point moves in the parameter space that gives the phase transition boundary.

It should be noted that the zero modes and the exceptional point are different concepts in both physical and mathematical senses. Consequently, the topological invariants based on the velocity field, the Chern number and winding number give different parameter-dependent invariants. The Chern number depends on $c$ because the energy gap is related to $c$ for quantum torus. When the energy band closes (exceptional point) the Chern number is ill-defined. Whereas the Euler characteristic associated with the zero modes is independent of $c$ because the Euler characteristic describes the global property of the submanifold even though the local behaviors of the velocity field depend on the shifting parameter $c$. It should be an interesting issue what physical phenomena emergence from these local and global properties of the velocity field in quantum states.

In summary, we propose the velocity field approach to capture the topological invariant of quantum states, which connects to the Euler characteristic by the Poincar\'{e}-Hopf theorem. We demonstrate the validity of this approach by the quantum torus model. The zero modes of the velocity field play the effective topological charges or defects to dominate the topological invariant of quantum states. We find that the different parameter-dependent of the topological invariants of quantum torus for the velocity field approach and the Chern number. We also propose the generalized winding number in terms of the velocity field for both Hermitian and non-Hermitian systems. This gives a connection between the zero mode and winding number in the velocity space.
These results not only provide a novel insight to topological invariants of quantum states and potential applications, but also inspire some fundamental issues what physical phenomena emergence associated with different topological invariants characterized by the velocity field and Chern number, especially what relationships between the local velocity field flow varying and the global topological invariant of the velocity field. The velocity field approach can be generalized to more generic and higher dimensional systems.

\begin{acknowledgments}
Authors thank the Grant of Scientific and Technological Projection of Guangdong province No: 2020B1212060030.
\end{acknowledgments}


\bibliography{apssamp}

\begin{thebibliography}{99}
\bibitem{Ramy} Ramy El-Ganainy, Konstantinos G. Makris, Mercedeh Khajavikhan,Ziad H.Musslimani, Stefan Rotter and Demetrios N.Christodoulides, Nature Phys. {\bf 14}, (2018) 11.
\bibitem{Kohei} Kohei Kawabata, Ken Shiozaki, Masahito Ueda, and Masatoshi Sato, Phys. Rev. {\bf X 9},(2019) 041015.

\bibitem{Gong} Zongping Gong, Yuto Ashida, Kohei Kawabata, Kazuaki Takasan, Sho Higashikawa, and Masahito Ueda, Phys. Rev. {\bf X 8},(2018) 031079.

\bibitem{Chiu} Ching-Kai Chiu, Jeffrey C.Y. Teo, Andreas P. Schnyder, Shinsei Ryu, {\it Rev. Mod. Phys.}, {\bf 88}, (2016) 035005.

\bibitem{Kane1} M.Z. Hasan, C. L. Kane, {\it Rev. Mod. Phys.} {\bf 82}, 3045 (2010).

\bibitem{Qi} Xiao-Liang Qi, Taylor L. Hughes, Shou-Cheng Zhang{\it Phys. Rev} {\bf B 78}, 195424 (2008).

\bibitem{Chruscinski} D. Chruscinski and A. Jamiolkowski, {\it Geometric Phase in classical and quantum mechanics}, Birkhauser Press; (2004).


\bibitem{Annan} Annan Fan, Guang-Yao Huang, Shi-Dong Liang, J. Phys. Commun. {\bf 4},  (2020) 115006.

\bibitem{Bohm}A. Bohm et al., {\it The Geometric Phase in Quantum Systems}, Springer, New York; (2003); Di Xiao, Ming-Che Chang, Qian Niu, {\it Rev. Mod. Phys.}, {\bf 82}, (2010) 1959.

\bibitem{Ghatak}Ananya Ghatak, and Tanmoy Das, J. Phys. CM {\bf 31}  (2019) 263001.

\bibitem{Liang} Shi-Dong Liang, Guang-Yao Huang, Phys. Rev. {\bf A 87},  (2013) 012118.

\bibitem{Zhang}Lin Zhang, Long Zhang, Sen Niu, and Xiong-Jun Liu, Sci. Bull. 63,  (2018) 1385.

\bibitem{Lee} T.E. Lee, Phys. Rev. Lett. {116}, (2016) 133903;  K. Esaki, M. Sato, K. Hasebe, and M. Kohmoto, Phys. Rev. {\bf B 84}, (2011) 205128.
\bibitem{Yao}Shunyu Yao and Zhong Wang,  Phys. Rev. Lett. {\bf 121} 086803 (2018); S. Yao, F. Song, and Z. Wang, Phys. Rev. Lett. {\bf 121}, (2018) 136802.

\bibitem{Yu} Yu Chen, and Hui Zhai, Phys. Rev. {\bf B 98}, 245130 (2018); T. Liu, Y.-R. Zhang, Q. Ai, Z. Gong, K. Kawabata, M. Ueda, and F. Nori, Phys. Rev. Lett. {\bf 122}, (2019) 076801.
\bibitem{Yin} Chuanhao Yin, Hui Jiang, Linhu Li, Rong Lu, and Shu Chen, Phys. Rev. {\bf A97}, (2018) 052115.


\bibitem{Fu} Huitao Shen, Bo Zhen, and Liang Fu, Phys. Rev. Lett. {\bf 120},  (2018) 146402.

\bibitem{Kane2} Jeffrey C. Y. Teo and C. L. Kane, Phys. Rev.{\bf B 82},  (2010) 115120.


\bibitem{Song} S. Lin, L. Jin, and Z. Song, Phys. Rev.{\bf B 99},  (2019) 165148; H. C. Wu, X. M. Yang, L. Jin ,
and Z. Song, Phys. Rev.{\bf B 102},  (2020) 161101(R); L. Jin and Z. Song, Phys. Rev.{\bf B 99},  (2019) 081103.

\bibitem{Eber} Eberhard Zeidler, Quantum Field Theory I: Basics in Mathematics and Physics, Springer, (2006); https://encyclopediaofmath.org/


\bibitem{Mermin}Neil W. Ashcroft and N. David Mermin, Solid state physics, Thomson Learning, Inc. (1976).






\end{thebibliography}

\end{document}